\documentclass[epj-spec]{svjour}
\usepackage{graphics}

\usepackage{amsmath,amssymb,graphicx,afterpage,here}
\usepackage{afterpage,subfigure,epsfig,psfig,rotating,minitoc}
\usepackage[english,activeacute]{babel}
\usepackage{ae}
\usepackage{dcolumn}% Align table columns on decimal point
\newcommand{\be}{\begin{equation}}
\newcommand{\ee}{\end{equation}}
\newcommand{\bea}{\begin{eqnarray}}
\newcommand{\eea}{\end{eqnarray}}

\newcommand{\bb}[1]{{\mbox{\boldmath $#1$}}}
\newcommand{\pd}{\partial}

\newcommand{\LB}[1]{\label{#1}}
\renewcommand{\vec}[1]{\mathbf{#1}}

\begin{document}
\title{Reentrant and Whirling Hexagons in Non-Boussinesq convection}
\author{ Santiago Madruga\inst{1}  \and Hermann Riecke\inst{2}  }
\institute{ Max-Planck-Institute for Physics of Complex Systems, D-01187
Dresden, Germany  \and Department of Engineering Sciences and Applied Mathematics,
Northwestern University, Evanston, IL 60208, USA }
\abstract{
We review recent computational results for hexagon patterns in
non-Boussinesq convection. For sufficiently strong dependence of the
fluid parameters on the temperature we find reentrance of steady
hexagons, i.e. while near onset the hexagon patterns become unstable 
to rolls as usually, they become again stable in the strongly
nonlinear regime. If the convection apparatus is rotated about a
vertical axis the transition from hexagons to rolls is replaced by a
Hopf bifurcation to whirling hexagons. For weak non-Boussinesq effects
they display defect chaos of the type described by the two-dimensional
complex Ginzburg-Landau equation. For stronger non-Boussinesq effects
the Hopf bifurcation becomes subcritical and localized bursting of the
whirling amplitude is found. In this regime the coupling of the
whirling amplitude to (small) deformations of the hexagon lattice
becomes important. For yet stronger non-Boussinesq effects this
coupling breaks up the hexagon lattice and strongly disordered states
characterized by whirling and lattice defects are obtained. 
} %end of abstract
%
%\pacs{47.20.Bp, 47.52.+j,5.45.Jn,47.54.+r}
\maketitle
\section{Introduction}

Rayleigh-B\'enard convection has been a paradigmatic system for
studies of spontaneously forming spatial or spatio-temporal patterns
(e.g. \cite{BoPe00}) far from equilibrium. In recent years exciting
results have been obtained for the stability and dynamics of
structures that are connected with roll convection. In particular, the
spiral-defect chaos obtained in convection of fluids with low Prandtl
number \cite{MoBo93,DePe94a,MoBo96,EgMe00,MaWo01,ChPa03} and domain
chaos \cite{ZhEc92,BoCa92,HuEc95,HuPe98} driven by the K\"uppers-Lortz
instability \cite{KuLo69} in rotating systems have attracted notable
interest. In most of these investigations the systems have been kept
close to the Oberbeck-Boussinesq approximation by minimizing the
dependence of the fluid parameters on the temperature  to avoid the
appearance of cellular or hexagonal structures. 

Variations of the fluid parameters with the temperature, i.e.
non-Oberbeck-Boussinesq effects, break the up-down symmetry and give
rise to hexagons at the onset of the convection. Weakly nonlinear
analysis shows that typically hexagons become unstable to rolls
further above threshold \cite{Bu67}. This scenario has been confirmed
in quite a number of experimental investigations \cite{BoPe00} and it
has often been assumed that in convection hexagonal patterns are
confined to the regime close to onset. Presumably due to this
assumption, hexagonal convection in strongly nonlinear non-Boussinesq
convection has not received as much attention.

Recently a couple of experiments \cite{AsSt96,RoSt02} showed that even
in the strongly nonlinear regime hexagonal convection patterns can be
observed stably. In one case hexagons were observed at relatively high
Rayleigh numbers ($\epsilon \equiv (R-R_c)/R_c \approx 3.5$) under
conditions in which the Oberbeck-Boussinesq approximation was quite
well satisfied. Correspondingly, hexagons with up- and with down-flow
in the center were observed to coexist in adjacent domains
\cite{AsSt96}. Subsequently, the linear stability of up- and
down-hexagons in this regime of Boussinesq-convection was confirmed by
a numerical stability analysis \cite{ClBu96,BuCl99a}. In the other
case a strongly non-Boussinesq situation was investigated and hexagons
were observed not only close to threshold but also again as reentrant
hexagons at higher Rayleigh numbers, $\epsilon = {\mathcal O}(1)$
\cite{RoSt02}. As the non-Boussinesq effects were increased the
intermediate $\epsilon$-range over which rolls were the preferred
planform shrank and eventually hexagons were found to dominate rolls
from directly at onset all the way to $\epsilon = {\mathcal O}(1)$.
The restabilization of hexagons at larger $\epsilon$ was attributed to
the high compressibility of the fluid; the experiment was performed
with SF$_6$ near its thermodynamical critical point \cite{RoSt02}.  

If the chiral symmetry of the convective apparatus is broken by 
rotating it about the vertical axis the K\"uppers-Lortz instability
\cite{KuLo69} leads to new complex dynamics. Above a critical
frequency rate domain chaos sets in, which is characterized by a
persistent switching between roll patches with different orientation.
In the non-Boussinesq case hexagons are steady near threshold. However
weakly non-linear theory predicts a secondary Hopf bifurcation to
oscillations in which the three modes forming the hexagonal lattice
oscillate with the same frequency but with a phase shift of $2\pi/3$
\cite{Sw84,So85}.

Here we review recent computational results for strongly nonlinear
non-Boussinesq convection \cite{MaRi07,MaRi06,MaRi06a} and illustrate
them further with additional data. The paper is organized as follows.
After outlining the basic equations for non-Boussinesq convection
(Section \ref{sec:basicequations}) we discuss reentrant hexagons in
fluids with high and to some extent also with low Prandtl number
(section \ref{sec:co2stability}). The reentrance is clearly
illustrated with numerical simulations in relatively large systems. In
Section \ref{sec:rotation} we address non-Boussinesq convection in
rotating systems. We review the transition to whirling hexagons in the
super- and the subcritical case, present new results for a
period-doubled state, and illustrate the mechanism underlying the
apparently persistent whirling chaos, which involves defects in the
hexagon lattice. Conclusions are given in Section
\ref{sec:conclusion}.

\section{Basic equations} \LB{sec:basicequations}

The basic equations that we use for the description of non-Boussinesq
convection have been discussed in detail previously
\cite{YoRi03b,MaRi06}. We therefore sketch here only a brief summary. 
We consider a horizontal fluid layer of thickness $h$, density $\rho$,
kinematic viscosity $\nu$, heat conductivity $\lambda$, thermal
diffusivity $\kappa$, and specific heat $c_p$. The system is heated
from below (at  temperature $T_1$) and cooled from above (at
temperature $T_2 < T_1$), and in the rotating case it rotates about a vertical axis with angular
velocity $\omega$, otherwise $\omega=0$. 

To render the governing equations and boundary conditions
dimensionless we choose the length $h$, the time $h^{2}/\kappa_0$, the
velocity $\kappa_0/d$, the pressure $\rho_0\nu_0 \kappa_0/h^{2}$, and
the temperature $T_s=\nu_0 \kappa_0/\alpha_0 g h^3$ as the respective
scales. The subscripted quantities refer to the respective values at
half depth of the fluid layer in the conductive state. The
non-dimensionalization gives rise to three dimensionless quantities:
the Prandtl number $Pr=\nu_0/\kappa_0$, the Rayleigh number
$R=\alpha_0 \Delta T g h^3/\nu_0 \kappa_0$, and the dimensionless
rotation rate  $\Omega=\omega\,h^2/\nu_0$.  Furthermore, we write the
equations in terms of the dimensionless momentum density $v_i=\rho h
u_i/\rho_0 \kappa_0$ instead of the velocities $u_i$. The
dimensionless form of the temperature $\hat T =T/T_s$, heat
conductivity $\hat \lambda =\lambda/\lambda_0$, density  $\hat \rho
=\rho/\rho_0$, kinematic viscosity $\hat \nu =\nu/\nu_0$, and specific
heat $\hat c_p =c_p/c_{p0}$ will be used in the  ensuing equations and
the hats dropped for simplicity. In dimensionless form the equations
for the momentum, mass conservation and heat are then given,
respectively, by 
\bea
\frac{1}{Pr}\left(\pd_tv_i+v_j\pd_j\left(\frac{v_i}{\rho}\right)\right)&=&-\pd_i
p  \LB{e:v} \\
&&+\delta_{i3}\left(1+\gamma_1(-2 z+\frac{\Theta}{R})\right)\Theta\nonumber \\
&&+\pd_j\left[\nu\rho\left(\pd_i(\frac{v_j}{\rho})+\pd_j(\frac{v_i}{\rho})\right)\right]
\nonumber \\
&&+2\Omega\,\epsilon_{ij3}v_j \nonumber \\
\pd_jv_j&=&0, \LB{e:cont}\\
\pd_t\Theta+\frac{v_j}{\rho}\pd_j\Theta
& =&\frac{1}{\rho
c_p}\pd_j(\lambda\pd_j\Theta)-\gamma_3\pd_z\Theta-\nonumber \\
&& R\frac{v_z}{\rho}(1+\gamma_3z).\LB{e:T}
\eea
with the dimensionless boundary conditions 
\bea
\vec{v}(x,y,z,t)=\Theta(x,y,z,t)=0  \mbox{ at } z= \pm \frac{1}{2}.\LB{e:bc}
\eea

Here $\Theta$ is the deviation
of the temperature field from the basic conductive profile, $\delta_{ij}$ is
the Kronecker delta, and $\epsilon_{ij3}$ the Levi-Civita tensor. Summation
over repeated indices is assumed.

In the rotating case, we assume small rotation rates and neglect the
centrifugal force term relative to gravity, i.e., we assume $\Omega^2
r <<g$. Notice that rotation breaks the chiral symmetry through  the
presence of the Coriolis term $2\bb{\Omega}\times \bb{u}$.

We consider the non-Boussinesq effects to be weak and retain in a
Taylor expansion of all material properties only the leading-order
temperature dependence {\it beyond} the OB approximation. For the
density this implies also a quadratic term with coefficient
$\gamma_1$. It contributes, however, only to the buoyancy term in
(\ref{e:v}); in all other expressions it would constitute only a
quadratic correction to the leading-order non-Boussinesq effect. Thus, the
remaining temperature dependence of the fluid parameters $\rho$,
$\nu$, $\lambda$, and $c_p$ in (\ref{e:v},\ref{e:cont},\ref{e:T}) is
taken to be linear \bea
\rho(\Theta)&=&1-\gamma_0(-z+\frac{\Theta}{R}),\LB{e:rhoTh}\\
\nu(\Theta)&=& 1+\gamma_2(-z+\frac{\Theta}{R}),\LB{e:nuTh}\\
\lambda(\Theta)&=&1+\gamma_3(-z+\frac{\Theta}{R}),\LB{e:lambdaTh}\\
c_p(\Theta)&=&1+\gamma_4(-z+\frac{\Theta}{R}).\LB{e:cpTh} \eea

The coefficients $\gamma_i$ give the difference of the respective
fluid properties across the layer. They depend therefore linearly on
the Rayleigh number, 
\bea
\gamma_i(\Delta T)=\gamma_i^{c}\,\left(\frac{R}{R_c} \right) 
=\gamma_i^{c} \, (1+\epsilon) ,
\eea
where $\gamma_i^{c}$ is the value of $\gamma_i$ at the onset of convection
and $\epsilon\equiv (R-R_c(\gamma_i^{c}))/R_c(\gamma_i^{c})$ is the
reduced Rayleigh number. 

The linear analysis yields the critical Rayleigh number $R_c$ as well
as the critical wavenumber $q_c$. Both depend on the non-Boussinesq-coefficients
$\gamma_i^{c}$ which in turn depend on $R_c$. Thus, in principle one
obtains an implicit equation for the $\gamma_i^{c}$. The shift in the
critical Rayleigh number away from the classical value $R_c=1708$ due
to the non-Boussinesq-effects is, however, quite small (less than 1 percent) and
therefore the resulting change in the $\gamma_i^{c}$ is negligible. In
this paper we therefore choose the $\gamma_i^{c}$ corresponding to
$R_c=1708$. 

Details of the numerical scheme used to determine the stability of
patterns in the strongly non-linear regime, and for the integration of
the Navier-Stokes equations can be found in reference \cite{MaRi06}.

\section{Reentrant hexagons} 
\LB{sec:co2stability}

The classic weakly nonlinear analysis for weakly non-Boussinesq
convection \cite{Bu67} predicts that not far above threshold the
hexagons become unstable to a mixed-mode, which in turn is also
unstable, resulting in a transition to roll convection. This
transition has been addressed in experimental studies in CO$_2$
\cite{BoBr91} and water \cite{PePa90,PaPe92}, respectively. In both
studies a transition to rolls was observed. The quantitative agreement
with the weakly nonlinear theory was, however, limited. Among the
reasons identified for the poor agreement were the boundaries, which
enhance the tendency towards rolls. 

In recent work on the stability of fully nonlinear hexagon solution of
the Navier-Stokes equations we have again addressed the transition
from hexagons to rolls \cite{MaRi06,MaRi07} and found that a further
cause for disagreement between weakly nonlinear theory and experiments
may be given by the fact that even for seemingly moderate
non-Boussinesq effects the transition from hexagons to rolls can be
shifted towards larger values of $\epsilon$ and can, in fact,
disappear altogether. 

Results of our stability calculation for a layer of water of height
$h=1.8\, {\rm mm}$ are shown as a function of the mean temperature
$T_0$ of the layer in  Fig.\ref{f:phase-dia}. The mean temperature is defined
as $T_0=(T_1+T_2)/2$,  where $T_1$ and $T_2$ are the temperatures at the
bottom and top plates respectively, and  $T_0$ is kept constant in the simulations of the Navier-Stokes and energy equations.
 Hexagons are stable
below and to the left of the solid line, whereas rolls are linearly
stable to hexagons above the dashed line. With decreasing mean
temperature (and with decreasing layer height) the critical
temperature for the onset of convection increases and with it the
variation of the fluid parameters across the layer. In addition, for
lower mean temperatures the temperature dependence of the density of
water develops a strong quadratic contribution. Thus, with decreasing
mean temperature the non-Boussinesq effects increase. This shifts the
line indicating the instability of hexagons to rolls to larger values
of $\epsilon$. More importantly yet, we found that the hexagons can
regain stability with respect to rolls if the Rayleigh number is
increased well beyond the weakly nonlinear regime. This line of
reentrance is shifted towards lower value of $\epsilon$ as the
non-Boussinesq effects become more pronounced and in fact merges
eventually with the lower stability limit of the hexagons at a
temperature $T_m$. Below $T_m$ the hexagons do not undergo any
instability with respect to rolls over the whole range of $\epsilon$
shown. 

\begin{figure}
\centering
\includegraphics[width=0.4\textwidth]{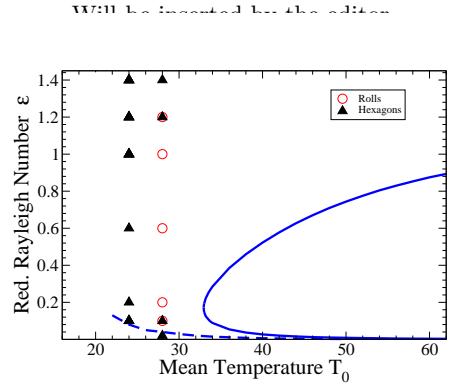}
\caption{(Color online). Phase diagram showing the results of simulations for a layer
of water with thickness $h=1.8\,{\rm mm}$ in a circular cell with diameter
$L=16\times 2\,\pi/q_c$. The simulations have been carried out for
$T_0=24\,^oC$ and $\,28\,^oC$ (for further parameters see
Figs.\ref{f:sim24},\ref{sim:water-to28}). Circles correspond to rolls
and triangles to hexagons.  The full and dashed line correspond to
the  stability limits for amplitude instabilities of hexagons and
rolls, respectively, as determined by the Galerkin calculation.
\LB{f:phase-dia}}
\end{figure}

More detailed stability analyses of the hexagons with periodic
boundary conditions using a general Galerkin-Floquet ansatz to capture
side-band instabilities show that the only relevant side-band
perturbations are long-wave and steady, as is the case in the weakly
nonlinear regime. The range of stable wavenumbers $q$ shifts towards
smaller values of $q$ as $\epsilon$ is increased. This reflects the
fact that even in the absence of non-Boussinesq effects hexagons can
become linearly stable relatively far from threshold, $\epsilon >2 $.
These hexagons are, however, stable with respect to side-band
instabilities only for relatively low values of $q$ \cite{ClBu96}. 

A seemingly similar reentrance of hexagon convection has been observed
by Roy and Steinberg in SF$_6$ near the thermodynamical critical point
\cite{RoSt02}. There it was argued that since the non-Boussinesq
effects in that system are not very large near onset the reentrance is
due to the large compressibility of the fluid in this parameter
regime. By assuming that the working fluid is incompressible, which is
an excellent approximation for water, our computations show that high
compressibility is not needed for reentrance. The significance of the
compressibility for the occurrence of the reentrance has also been
called into question in \cite{Ah05,OhOr04}. Moreover, the wavenumber
of the reentrant hexagons of \cite{RoSt02} does not decrease with
increasing $\epsilon$, but stays quite close to $q_c$, whereas a
characteristic feature of the reentrant hexagons found in our
computation is that the wavenumber is significantly lower than $q_c$
(cf. \cite{ClBu96}). 

With respect to the experiments in water \cite{PaPe92} our stability
computations show that even in that system reentrance should have
been accessible just slightly above the Rayleigh numbers investigated
in those experiments ($0 < \epsilon < 0.14$). For the experiment in
CO$_2$ \cite{BoBr91} we find that the parameters correspond to the
regime in which hexagons do not become unstable to rolls at all up to
$\epsilon=1$, similar to the situation shown in Fig.\ref{f:phase-dia}
for water for $T_0<T_m$. 

To address the influence of the side walls we have performed numerical
simulations of the Navier-Stokes equations in which we mimic the
circular sidewalls by applying a strong radial subcritical ramp in the
Rayleigh number that suppress any convection outside a certain radius 
\cite{DePe94a}. The system size is chosen to be twice as large as in
\cite{MaRi07}, i.e. here $L=16\times 2\pi/q_c$. Random initial
conditions are used. Fig. \ref{f:sim24} shows snapshots of simulations
for a mean temperature of $T_0=24^\circ C$. Consistent with the phase
diagram shown in Fig.\ref{f:phase-dia} no transition from hexagons to
rolls is found. As expected from the shift of the stable wavenumber
band towards smaller value of $q$ the hexagon wavenumber is decreasing
with increasing $\epsilon$. Note that the defects that cause the
change in wavenumber do not trigger a transition to rolls.

\begin{center}
\begin{figure}
\centering
\includegraphics[width=0.48\textwidth]{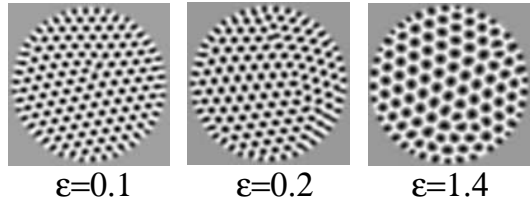}
\caption{ Series of snapshots in a circular cell of water
of thickness $h=1.8\,{\rm mm}$ and mean temperature $T_0=24\,^oC$. The diameter of 
the cell is $L=16\times2\pi/q_c$,
and the snapshots correspond to an integration time of $600\,t_v$. 
The corresponding non-Boussinesq coefficients are $\gamma_0^c=0.0039$, $\gamma_1^c=0.3326$,
$\gamma_2^c=-0.3641$, $\gamma_3^c=0.0467$, $\gamma_4^c=-0.0026$, and the Prandtl number $\text{Pr}=6.3$.
\LB{f:sim24}} 
\end{figure}
\end{center}

For larger  mean temperature, $T_0=28^oC$, the non-Boussinesq effects
are weaker.  As a consequence, for intermediate values of $\epsilon$
rolls invade the hexagon pattern from the side walls similar to the
experiments of \cite{PaPe92,BoBr91} and  for values of $\epsilon$  in
the range $0.6<\epsilon <0.1$ disordered roll patterns are obtained,
which are characteristically perpendicular to the sidewalls
(Fig.\ref{sim:water-to28}). For $\epsilon=1.2$ the hexagons take over
again and we find a new interesting scenario: coexistence between
rolls and reentrant hexagons. This state was not obtained in our
previous simulations of water due to the small aspect ratio used
\cite{MaRi06} nor in simulations of reaction diffusion system
\cite{VeWi92}. At $\epsilon=1.4$ the non-Boussinesq effects become
strong enough to remove all the rolls and the reentrant hexagons take
up the whole convective cell. 

The results of the simulations are summarized in Fig.\ref{f:phase-dia}
with triangles denoting hexagons and circles marking roll states.
Compared to Fig.9 in \cite{MaRi06} the regime of disordered roll
patterns is shifted towards larger $T_0$, i.e towards weaker non-
Boussinesq effects. This is due to the larger convection cell used in
the present simulations. This reduces the influence of the sidewalls.
They enhance the nucleation of rolls, which then can invade the
hexagonal pattern in the bulk.  

\begin{figure}[tp!]
\centering
\includegraphics[width=0.48\textwidth]{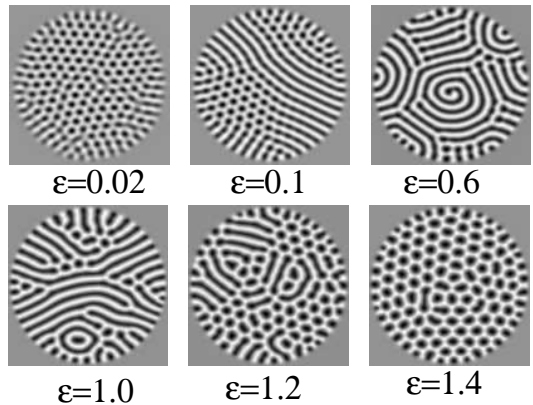}
\caption{Sequence of snapshots for $T_0=28\,^oC$ in a circular cell of water
of thickness $h=1.8\,{\rm mm}$. The diameter of the cell is $L=16\times2\pi/q_c$,
and the snapshots correspond to an integration time of $300\,t_v$. 
The corresponding non-Boussinesq coefficients are $\gamma_0^c=0.0036$, $\gamma_1^c=0.2122$,
$\gamma_2^c=-0.2725$, $\gamma_3^c=0.0352$, $\gamma_4^c=-0.0013$, and the Prandtl number $\text{Pr}=5.7$.
\LB{sim:water-to28}} 
\end{figure}

\section{Convection with rotation}
\LB{sec:rotation}

If the whole system is rotated about a vertical axis the Coriolis
force breaks the chiral symmetry, which transforms the steady
bifurcation off the hexagons to the unstable mixed mode into a Hopf
bifurcation to whirling hexagons \cite{Sw84,So85}. In
\cite{EcRi00a,EcRi00b} the complex Ginzburg-Landau equation describing
spatial modulations of the complex oscillation amplitude ${\mathcal
H}$ of the whirling hexagons has been derived from three coupled
Ginzburg-Landau equations that provide the weakly nonlinear
description of the hexagons,
\begin{eqnarray}
\partial_T{\cal H}&=&a{\cal H}+d\nabla^2{\cal H}-c{\cal H}|{\cal
H}|^2.
\label{e:cgl} 
\end{eqnarray}

In principle, this CGL is still coupled to the phase modes of the
underlying hexagon pattern. However, at the band center $q_c$ the
coupling coefficients vanish, irrespective of the coefficients of the
three coupled Ginzburg-Landau equations. Moreover, the imaginary parts
of the  coefficients $d$ and $c$ are always in the regime of
bistability of the homogeneous oscillations and defect chaos
\cite{ChMa96}.

Recently, we have investigated this Hopf bifurcation within the fully
nonlinear Navier-Stokes equations (\ref{e:v},\ref{e:cont},\ref{e:T})
and have in particular obtained the coefficients of the CGL (\ref{e:cgl}) 
for realistic fluid parameters \cite{MaRi06a}.  To this end we have
performed a numerical linear stability analysis of the steady hexagons
employing a Galerkin expansion of the velocity and temperature fields
for the Floquet analysis. 

 The linear coefficient
$a\equiv a_r+ia_i$ is readily obtained from the growth rate and
frequency of the perturbation associated with the Hopf bifurcation.
The calculation of the diffusion and dispersion coefficient $d\equiv
d_r+i\,d_i$ is a bit more involved. It captures the dependence of the
complex growth rate $\sigma$ of the oscillatory Hopf mode on slow
spatial modulations of its amplitude,  $\sigma=-s^2\,d$, where $s$ is
the wavenumber of the modulation, i.e. the Floquet exponent. Note,
that the dependence of the growth rate of the translation mode (rather
than the Hopf mode) on the Floquet parameter would yield the phase
diffusion coefficients. We choose a discrete number of modulation
wavevectors $\vec{s}$ typically in the range $0.01 \le |\vec{s}|/q_c
\le 1$ for the computations and obtain $d$ by a fit of $\sigma$ to the
equation $\sigma=-s^2\,d$. Fig.\ref{f:drdi-w-sup} shows the ratio
$d_i/d_r$ for $T_0=14^\circ C$ and $h=4.92\,{\rm mm}$ as a function of
the deviation of wavenumber of the underlying hexagons from the
critical wavenumber, $\Delta q\equiv q-q_c$. Within the coupled
Ginzburg-Landau equations $d_i/d_r$ vanishes at the bandcenter,
$\Delta q=0$, and is quadratic in $\Delta q$. The fully nonlinear
computations of the Navier-Stokes equations confirm this behavior
quite well.

\begin{figure}[t!]
\centering
\includegraphics[width=0.35\textwidth,angle=0]{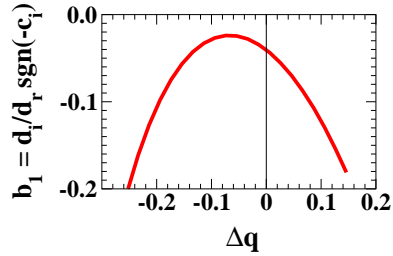}
\caption{(Color online). Dependence of the ratio  $d_i/d_r$ on the wavenumber
$\Delta q\equiv q-q_c$. $T_0=14^oC$, $\Omega=65$, non-Boussinesq coefficients 
$\gamma_0^c=0.0009$, $\gamma_1^c=0.2937$,
$\gamma_2^c=-0.1681$, $\gamma_3^c=0.0215$, $\gamma_4^c=-0.0022$, and
Prandtl number $\text{Pr}=8.2$.}
\LB{f:drdi-w-sup}
\end{figure}

For the calculation of the non-linear coefficient $c\equiv c_r+ic_i$
we perform direct simulations of the Navier-Stokes equations
(\ref{e:v}-\ref{e:bc}) in a small box of size  $L=2\times 2\,\pi/q_c$
to avoid side-band perturbations and spatial modulations of the
oscillation amplitude. Therefore, from the real part
$dH/dt=a_r|{\mathcal H}|+c_r|\mathcal{H}|^3$ we extract $c_r$ by
fitting the time derivative of the amplitude of the Hopf mode as a
function of its amplitude (notice that we extracted $a_r$ in the
previous step through a stability linear analysis). Finally, from the
imaginary part $\omega=a_i+c_i|\mathcal{H}|^2$ we obtain $c_i$ by
fitting the oscillatory frequency of the Hopf mode as a function of
its amplitude. 

For $T_0=14^\circ C$ and $h=4.92{\rm mm}$ with $\Omega=65$ we obtain
coefficients that match closely the expectations based on the weakly
nonlinear theory provided by the three coupled Ginzburg-Landau
equations. In particular, we confirm that the system should exhibit
bistability between homogeneous oscillations and defect chaos. Direct
simulations in sufficiently large system confirmed this prediction.
When starting with ordered hexagons and small perturbations that break
the chiral symmetry, i.e. the three dominant wavevectors making up the
hexagon pattern are given slightly different amplitudes, we obtain
stable homogeneous oscillations, i.e. all hexagons whirl in phase with
each other. To obtain the defect-chaos, defects need to be introduced
into the complex oscillation amplitude. Note, however, that
preferably there should be no defects in the underlying hexagon
pattern. We therefore start with initial conditions in which the
complex amplitude exhibits phase winding and its phase changes by
$2\pi$ across the system. In our simulations we find that this
phase-modulated state is linearly unstable to side-band perturbations
of the complex oscillation amplitudes which lead to defects in that
amplitude. Through their wavenumber selection the defects induce the
creation of further defects, without any significant deformations of
the hexagon lattice. 

For somewhat stronger non-Boussinesq effects ($T_0=12^\circ C$ with
$4.92{\rm mm}$) we find that the Hopf bifurcation becomes subcritical.
Since within the three coupled Ginzburg-Landau equations the Hopf
bifurcation is always supercritical, this shows that the Hopf
bifurcation occurs now beyond their regime of validity. Nevertheless,
near the parameter values for which the bifurcation changes direction
the whirling hexagons can still be described by a complex
Ginzburg-Landau equation if a quintic saturating term $-g|{\mathcal
H}^4{\mathcal H}$ is retained. Following the same procedure as before
we determine the coefficients of this quintic equation and obtain
$d=22.4+0.389i,\,c=-12.96+84.7i,\,g=42.53+17.20i$. The dependence of
$d_i/d_r$ on the hexagon wavenumber is shown in Fig.\ref{f:drdi-w}. In
contrast to the case $T_0=14^\circ C$ shown in Fig.\ref{f:drdi-w-sup},
the dependence is now essentially linear near $q_c$, reflecting again
the fact that the Hopf bifurcation occurs at larger amplitudes for
which the coupled Ginzburg-Landau equations are no longer valid. The
inset of Fig.\ref{f:drdi-w} shows that the dependence of $\omega$ with
respect to $\mathcal{H}^2$ is almost linear, and thus the quintic
contribution $g_i |{\mathcal H}|^4$ to the frequency is very small in
this regime. 

\begin{figure}[t!]
\centering
\includegraphics[width=0.4\textwidth,angle=0]{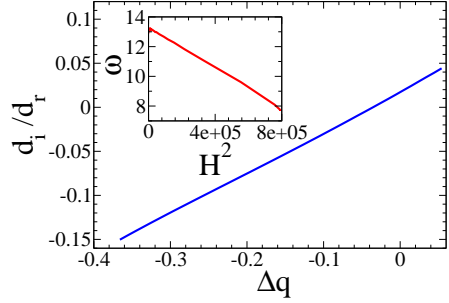}
\caption{Dependence of the ratio  $d_i/d_r$ on the wavenumber 
$\Delta q\equiv q-q_c$ (main figure), and dependence of the frequency 
of the Hopf mode on its amplitude (inset). The ratio $d_i/d_r$ is extracted 
from linear stability calculations and the frequency from numerical simulations 
in a small box of size $L=2\times 2\,\pi/q_c$, with $T_0=12^oC$, $\Omega=65$, 
non-Boussinesq coefficients 
$\gamma_0^c=0.00100$, $\gamma_1^c=0.4885$,
$\gamma_2^c=-0.2281$, $\gamma_3^c=0.0287$, $\gamma_4^c=-0.00320$, and
Prandtl number $\text{Pr}=8.7$.}
\LB{f:drdi-w}
\end{figure}

Direct simulations of the Navier-Stokes equations reveal that in this
regime typical states exhibit localized unsteady bursts in the
oscillation amplitude as shown in Fig.\ref{f:locstc}. The white lines
give the hexagonal cells, showing the deformations due to the
whirling, and the magnitude $|{\mathcal H}|$ of the oscillation
amplitude is shown in the color scale (red: maximal, blue: zero). We
extract the complex oscillation amplitude ${\mathcal H}$ from the
simulation by demodulation with respect to the wavenumber of the
hexagons and the dominant frequency of the oscillation. To do so, we
first evaluate the Fourier transform of the time series given by the
amplitude at each point of the numerical grid, and then shift the
resulting temporal Fourier spectrum by the Hopf frequency. In this way
we capture the slow time evolution. To obtain the slow spatial
modulations, we compute the spatial Fourier transform of the shifted
temporal Fourier transform, shift it such that one of the six dominant
peaks making up the hexagonal lattice is moved to the origin, and
remove the spectral content beyond a radius of $\sqrt{3}\,q_c/2$. 

\begin{figure}[t!]
\centering
\includegraphics[width=0.35\textwidth,angle=0]{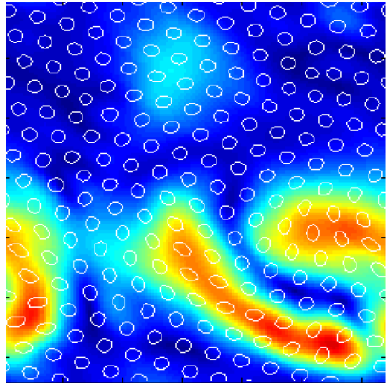}
\caption{(Color online). Localized bursting of the oscillation amplitude. Color gives
magnitude of oscillation amplitude (red: maximal, blue: zero). White
lines indicate underlying hexagon pattern and show the deformation of
the hexagon cells by the whirling.}
\LB{f:locstc}
\end{figure}

To assess the validity of the quintic complex Ginzburg-Landau equation
we simulate it numerically using the parameters obtained from the
simulations of the Navier-Stokes equation. We find that using the same
system size the quintic CGL does not exhibit the bursting found in the
Navier-Stokes simulations since the homogeneous oscillations become
unstable to side-band perturbations only for smaller wavenumbers. If,
however, the system size is chosen larger, similar bursting is
obtained, as shown in Fig.\ref{f:cgl-sub}. A more careful analysis
shows that in this regime the coupling to the phase equations for the
deformations of the underlying hexagons cannot be ignored
\cite{MaRi06a}. These deformations represent changes in the local
wavenumber of the hexagons. As our numerical stability analysis of the
steady hexagons shows, in this regime the growth rate of the Hopf mode
increases significantly with decreasing hexagon wavenumber. Thus, due
to the deformation of the hexagons by the whirling the local growth 
rate within a burst is enhanced and allows this burst to persist. 

\begin{figure}[t!]
\centering
\includegraphics[width=0.35\textwidth,angle=0]{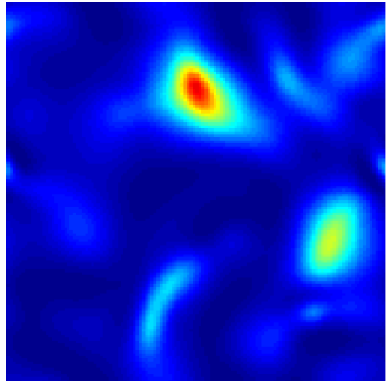}
\caption{(Color online). Localized bursting of the oscillation amplitude in the
quintic complex Ginzburg-Landau equation. Parameters correspond to
those of Fig.\ref{f:locstc} except for a larger system size.}
\LB{f:cgl-sub}
\end{figure}

\begin{figure}[tp!]
\centering
\includegraphics[width=0.4\textwidth,angle=0]{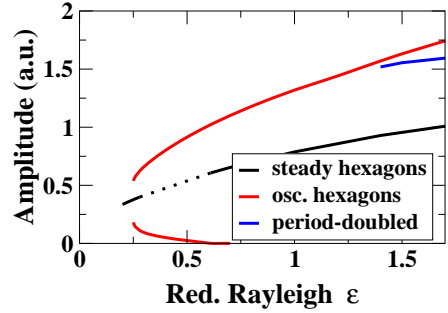}
\caption{(Color online). Bifurcation diagram of non-Boussinesq convection with
rotation for the subcritical case with  $T_0=12^\circ C$ and $\Omega=65$.
Blue line corresponds to the minimal oscillation amplitude of the
period-doubled orbit, the maximal oscillation amplitude is marked by
the red line.   } \LB{f:bif-rot-sub}
\end{figure}

Increasing the Rayleigh number, the steady hexagons again become
stable, as in the non-rotating case discussed above. The corresponding
bifurcation diagram is shown in Fig.\ref{f:bif-rot-sub}. Nevertheless,
for suitable initial conditions we still find irregular bursting of
the whirling activity for $\epsilon=0.7$, but not for $\epsilon=0.9$.
Two snapshots depicting whirling localized within domains of varying
sizes are shown in Fig.\ref{f:whirl-eps0.7}. Compared to the bursting
obtained for $\epsilon=0.5$ \cite{MaRi06a}, the fluctuations in the
whirling activity, $I(t)=N^{-1}\int_{|{\mathcal H}|>0.5{|\mathcal
H}|_{max}} |{\mathcal H}|\,dxdy$ with $N$ denoting the temporal mean
of the integral, are much smaller (Fig.\ref{f:whirl-time}).
Nevertheless, persistently shifting domains of whirling hexagons are
still apparent in Fig.\ref{f:whirl-eps0.7}. 

\begin{center}
\begin{figure}
\centering
\includegraphics[width=0.5\textwidth,angle=0]{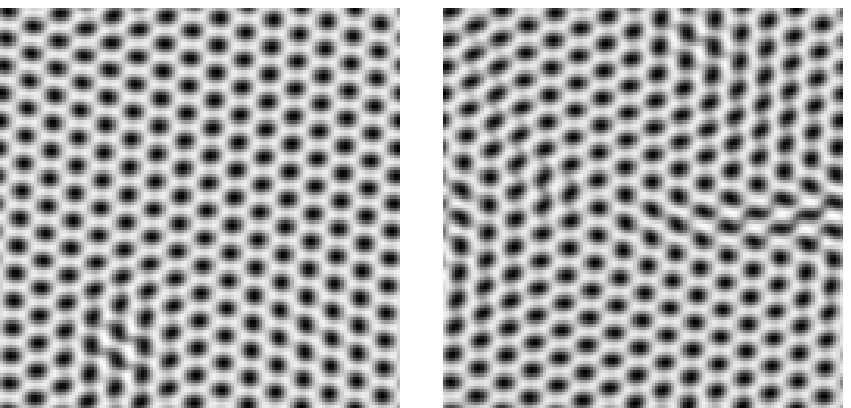}
\caption{Bursting state in water for $T_0=12^\circ C$, $\Omega=65$ and
$\epsilon=0.7$. The snapshots come from simulations of Eqs. (\ref{e:v}-\ref{e:bc}) 
in a cell of size $L=16\times2\pi/q_c$ using random
initial conditions. Left snapshot corresponds to time $t=663\,t_v $ and right $t=699\,t_v$.
} \LB{f:whirl-eps0.7}
\end{figure}
\end{center}

\begin{center}
\begin{figure}
\centering
\includegraphics[width=0.4\textwidth,angle=0]{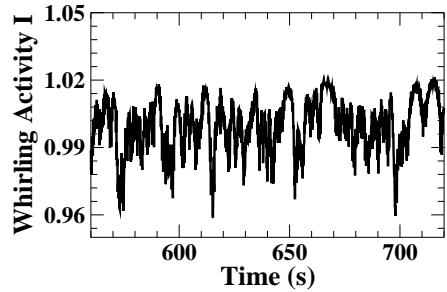}
\caption{Temporal evolution of the whirling activity  
$I(t)=N^{-1}\int_{|{\mathcal H}|>0.5{|\mathcal H}|_{max}}|{\mathcal H}|\,dxdy$ 
of the bursting state (see Fig.\ref{f:whirl-eps0.7}). 
} \LB{f:whirl-time}
\end{figure}
\end{center}

Increasing $\epsilon$ further, we find a period-doubling bifurcation
at $\epsilon \simeq 1.3$ as shown in Fig.\ref{f:bif-rot-sub}, which
gives the larger (red) and smaller (blue) peak amplitude of the
oscillations. Fig.\ref{f:double} presents a time series of
the amplitude of one of the Fourier modes that forms the hexagonal
lattice for $\epsilon=1.5$. The time series is obtained from the
simulation of  equations $(1-4)$ in a  small cell of size
$L=2\times2\,\pi/q_c$. Whether there are any further period doubling
events for larger $\epsilon$ remains to be investigated. So far we
have obtained this state only in simulations of small systems in which
side-band perturbations of the oscillation mode as well as the hexagon
pattern are negligible.

\begin{figure}[tp!]
\centering
\includegraphics[width=0.35\textwidth,angle=0]{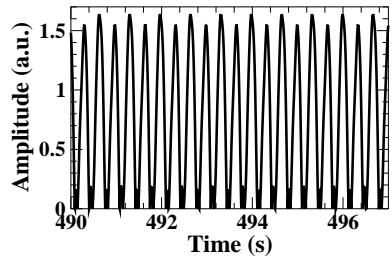}
\caption{Period doubling after a Hopf bifurcation in the amplitude
of one of the three modes that forms a hexagonal lattice. The simulations are
made in  a small cell of size of $L=2\times2\,\pi/q_c$ with $T_0=12$, $\epsilon=1.5$, and $\Omega=65$.}  
\LB{f:double}
\end{figure}

As the non-Boussinesq effects are increased yet further
\cite{YoRi03b}, the range over which the steady hexagons are unstable
to the whirling becomes smaller and eventually the steady hexagons are
linearly stable over the whole range of Rayleigh numbers considered,
$0\le \epsilon \le 1.2$. However, since the Hopf bifurcation is
subcritical in this regime solutions with whirling hexagons still
exist. In fact, the whirling activity, which does deform the
underlying hexagon pattern \cite{EcRi00a,EcRi00b,MaPe04}, 
is so strong that it breaks up the hexagonal lattice, introducing many
defects. In weakly disordered hexagon patterns the dominant defects
are typically penta-hepta defects which consist of two adjacent
convection cells that have 5 and 7 immediate neighbors, respectively.
The state resulting from the strong whirling activity is, however,
considerably more disordered and exhibits not only such cells with
coordination numbers 5 and 7, but also a small number of cells with
coordination numbers 4 and 8, respectively \cite{YoRi03b}.

For intermediate Rayleigh numbers the spatio-temporally chaotic
whirling activity arises only transiently. During that transient one
can qualitatively discern a mutual reinforcement of the defect
formation and the whirling activity. This is illustrated in the two
snapshots shown in Fig.\ref{f:defwhirl}. Whirling activity emanates
from the defects in the hexagonal lattice (marked by a red and a blue
circle). It in turn tends to induce other defects, see right panel.
For the parameters chosen in Fig.\ref{f:defwhirl} the whirling
emanating from the defects is not strong enough and the various
defects in the lattice eventually annihilate each other and the
pattern becomes steady. If the mutual reinforcement is strong enough,
however, persistent activity can arise. For $\epsilon=1$ the whirling
invades the steady state and in systems of size $L=16\times 2\pi/q$
with $q=4.5$ we find that the chaotic activity persists as long as we
could simulate the system (more than 300 vertical thermal diffusion
times, which corresponds to more than 5 hours physical time).

\begin{figure}[tp!]
\centering
\epsfxsize=4.1cm\epsfbox{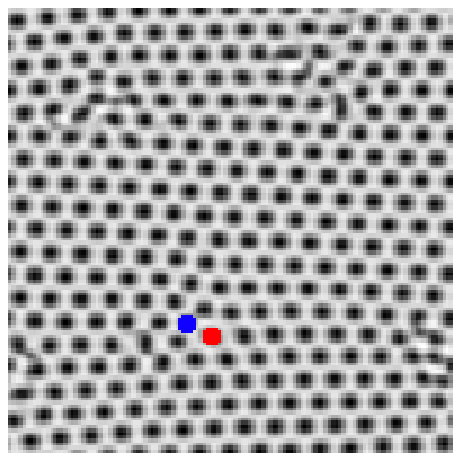}
\epsfxsize=4.1cm\epsfbox{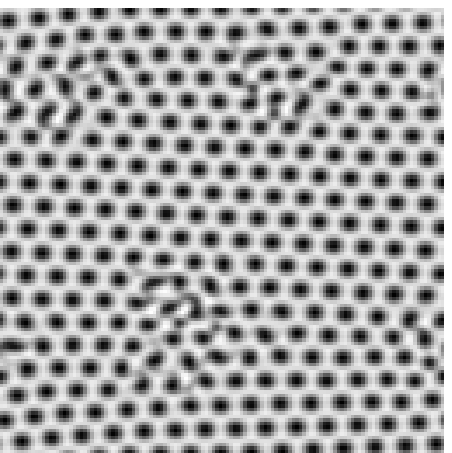} 
\caption{(Color online). Interplay between whirling activity and defect formation.
Two snapshots for $\epsilon=0.87$ and $h=4.6{\rm mm}$, $T_0=12^\circ C$
$\Omega=65$. Penta-hepta defects (one is marked by blue and red circle) 
induce whirling and whirling tends to induce defects. }.  
\LB{f:defwhirl}
\end{figure}

\section{Conclusion}
\LB{sec:conclusion}

In this paper we have elaborated on some recent results for hexagon
convection in non-Boussinesq convection \cite{MaRi06,MaRi06a,MaRi07}.
In the non-rotating case we focused on the reentrance of hexagons as
the Rayleigh number is increased. We contrasted the reentrance found
in our computations for water with that found in experiments using a
layer of SF$_6$. The reentrant hexagons obtained through the mechanism
elucidated with our computations generically are expected to inherit
the low wavenumber that is characteristic of Boussinesq hexagons
\cite{ClBu96}. In contrast, the reentrant hexagons found in the SF$_6$
experiments have a wavenumber close to $q_c$ \cite{RoSt02}, suggesting
that their reentrance is due to a different mechanism.

For stronger non-Boussinesq effects the hexagons do not exhibit any
linear instability to rolls (or the mixed mode) and, in fact, even
defects arising through side-band instabilities of the hexagons do not
necessarily lead to a transition to rolls. Instead the pattern can
heal to a regular hexagon pattern with lower wavenumber. In convection
in gases, which have a low Prandtl number, the side-band instabilities
limiting the stable wavenumber band can be different at the low-$q$
and the high-$q$ side, which leads to transients exhibiting different
degrees of disorder \cite{MaRi07}.

Similar looking reentrant hexagons have also been reported in a
chemical model system, the two-dimensional Brusselator system, for
which a series of numerical simulations were performed \cite{VeWi92}.
The initial Turing instability off the homogeneous state yields
hexagon patterns. With increasing control parameter these hexagons
lose their stability to stripe-like patterns. For yet larger values of
the control parameter the hexagons regain their stability. In contrast
to the behavior found in convection in this chemical model the
hexagons near onset are up-hexagons ($\pi$-hexagons) while the
reentrant hexagons are down-hexagons ($0$-hexagons). These results are
modeled with an amplitude equation whose quadratic coefficient depends
on the control parameter. Such a model has also has been used for
hexagons in magnetic fluids \cite{FrEn01} and in our previous
calculations for water \cite{MaRi06}. 

For rotating systems we discussed the whirling hexagons that arise
from a Hopf bifurcation that takes the place of the steady bifurcation
to a mixed mode in the non-rotating case. In the weakly non-Boussinesq
case this transition occurs for very small convection amplitudes and
the Hopf bifurcation is supercritical, as predicted by computations
based on three coupled Ginzburg-Landau equations describing weakly
nonlinear hexagon patterns with rotation \cite{EcRi00a}. It turns out
that to leading order the oscillations can be described by the
two-dimensional complex Ginzburg-Landau equation. Moreover, the
resulting oscillations (whirling hexagons) generically arise in the
regime in which defect chaos persists despite the existence of stable
plane wave and homogeneously oscillating solutions. For stronger
non-Boussinesq effects the bifurcation becomes subcritical and we find
localized bursting in the whirling amplitude and also whirling
activity that is strong enough to tear the hexagonal lattice apart.

This paper is dedicated to the memory of Carlos P\'erez-Garc\'{\i}a, who was
SM's thesis advisor and who introduced him to the field of
pattern formation. SM is grateful for years of friendship and stimulating
discussions.

We thank Werner Pesch for providing us with the numerical codes used in this
project and for ongoing support. 

This research was supported by NSF
(DMS-9804673), by the Department of Energy (DE-FG02-92ER14303), and by EU
under grant MRTN-CT-2004-005728 (SM).


\begin{thebibliography}{10}

\bibitem{BoPe00}
E.~Bodenschatz, W.~Pesch, and G.~Ahlers.
\newblock {\em Ann. Rev. Fluid Mech.}, 32:709--778, 2000.

\bibitem{MoBo93}
S.~W. Morris, E.~Bodenschatz, D.S. Cannell, and G.~Ahlers.
\newblock {\em Phys. Rev. Lett.}, 71:2026, 1993.

\bibitem{DePe94a}
W.~Decker, W.~Pesch, and A.~Weber.
\newblock {\em Phys. Rev. Lett.}, 73:648, 1994.

\bibitem{MoBo96}
S.~W. Morris, E.~Bodenschatz, D.~S. Cannell, and G.~Ahlers.
\newblock {\em Physica D}, 97:164, 1996.

\bibitem{EgMe00}
D.~A. Egolf, I.~V. Melnikov, W.~Pesch, and R.~E. Ecke.
\newblock {\em Nature}, 404:733--736, 2000.

\bibitem{MaWo01}
R.~G. Matley, W.~Y. Wong, M.~S. Thurlow, P.~G. Lucas, M.~J. Lees, O.~J.
  Griffiths, and A.~L. Woodcraft.
\newblock {\em Phys. Rev. E}, 63:045301, 2001.

\bibitem{ChPa03}
K.-H. Chiam, M.~R. Paul, M.~C. Cross, and H.~S. Greenside.
\newblock {\em Phys. Rev. E}, 67:056206, 2003.

\bibitem{ZhEc92}
F.~Zhong and R.E. Ecke.
\newblock {\em Chaos}, 2:163, 1992.

\bibitem{BoCa92}
E.~Bodenschatz, D.S. Cannell, J.~R. DeBruyn, R.~Ecke, Y.C. Hu, K.~Lerman, and
  G.~Ahlers.
\newblock {\em Physica D}, 61:77, 1992.

\bibitem{HuEc95}
Y.~Hu, R.E. Ecke, and G.~Ahlers.
\newblock {\em Phys. Rev. Lett.}, 74:5040, 1995.

\bibitem{HuPe98}
Y.~Hu, W.~Pesch, G.~Ahlers, and R.~E. Ecke.
\newblock {\em Phys. Rev. E}, 58:5821, 1998.

\bibitem{KuLo69}
G.~{K\"uppers} and D.~Lortz.
\newblock {\em J. Fluid Mech.}, 35:609, 1969.

\bibitem{Bu67}
F.~H. Busse.
\newblock {\em J. Fluid Mech.}, 30:625, 1967.

\bibitem{AsSt96}
Michel Assenheimer and Victor Steinberg.
\newblock {\em Phys. Rev. Lett.}, 76:756--759, 1996.

\bibitem{RoSt02}
A.~Roy and V.~Steinberg.
\newblock {\em Phys. Rev. Lett.}, 88:244503, 2002.

\bibitem{ClBu96}
R.~M. Clever and F.~H. Busse.
\newblock {\em Phys. Rev. E}, 53:R2037--R2040, 1996.

\bibitem{BuCl99a}
F.~H. Busse, R.~M. Clever, and E.~Grote.
\newblock {\em Chaos Solitons Fractals}, 10(4-5):753--760, 1999.

\bibitem{Sw84}
J.~W. Swift.
\newblock In {\em Contemporary Mathematics Vol. 28}, page 435, Providence,
  1984. American Mathematical Society.

\bibitem{So85}
A.~M. Soward.
\newblock {\em Physica D}, 14:227--241, 1985.

\bibitem{MaRi07}
S.~Madruga and H.~Riecke.
\newblock {\em To appear in: Phys. Rev. E}, 2007.

\bibitem{MaRi06}
S.~Madruga, H.~Riecke, and W.~Pesch.
\newblock {\em J. Fluid Mech.}, 548:341--360, 2006.

\bibitem{MaRi06a}
S.~Madruga, H.~Riecke, and W.~Pesch.
\newblock {\em Phys. Rev. Lett.}, 96:074501, 2006.

\bibitem{YoRi03b}
Y.-N. Young, H.~Riecke, and W.~Pesch.
\newblock {\em New J. Phys.}, 5:135, 2003.

\bibitem{BoBr91}
E.~Bodenschatz, J.~R. deBruyn, G.~Ahlers, and D.S. Cannell.
\newblock {\em Phys. Rev. Lett.}, 67:3078, 1991.

\bibitem{PaPe92}
E.~Pampaloni, C.~{P\'erez-Gar\'{\i}a}, L.~Albavetti, and S.~Ciliberto.
\newblock {\em J. Fluid Mech.}, 234:393--416, 1992.

\bibitem{PePa90}
C.~P\'erez-Garc\'{\i}a, E.~Pampaloni, and S.~Ciliberto.
\newblock {\em Europhys. Lett.}, 12(1):51--55, 1990.

\bibitem{Ah05}
G.~Ahlers.
\newblock In I.~Mutabazi, Jose~E. Wesfreid, and E.~Guyon, editors, {\em
  Dynamics of spatio-temporal cellular structures - {Henri} {B\'enard}
  centenary review}, Springer Tracts in Modern Physics. Springer, 2005.

\bibitem{OhOr04}
J.~Oh, J.~{Ortiz de Z\'arate}, J.~V. Sengers, and G.~Ahlers.
\newblock {\em Phys. Rev. E}, 69:021106, 2004.

\bibitem{VeWi92}
G.~Dewel J.~Verdasca, A. de~Wit and P.~Borkmans.
\newblock {\em Phys. Lett. A}, 168:194--198, 1992.

\bibitem{EcRi00a}
B.~Echebarria and H.~Riecke.
\newblock {\em Phys. Rev. Lett.}, 84:4838, 2000.

\bibitem{EcRi00b}
B.~Echebarria and H.~Riecke.
\newblock {\em Physica D}, 143:187, 2000.

\bibitem{ChMa96}
H.~{Chat\'{e}} and P.~Manneville.
\newblock {\em Physica A}, 224:348, 1996.

\bibitem{MaPe04}
S.~Madruga and C.~P\'erez-Garc\'{\i}a.
\newblock {\em Int. J. Bifurcation Chaos}, 14:107--117, 2004.

\bibitem{FrEn01}
R.~Friedrichs and A.~Engel.
\newblock {\em Phys. Rev. E}, 6402:021406, 2001.

\end{thebibliography}
\end{document}